\documentclass[aps,prl,reprint]{revtex4-1}
\usepackage[pdftex]{graphicx}% Include figure files
\usepackage{amssymb,amsfonts,amsmath}
\usepackage{color}
\usepackage[usenames,dvipsnames]{xcolor}
\usepackage{sidecap}

\begin{document}

\title{The free energy of grain boundaries from atomistic computer simulation}

\author{Saswati Ganguly}
\affiliation{Institut f\"ur Theoretische Physik II: Weiche Materie, 
Heinrich-Heine-Universit\"at D\"usseldorf, Universit\"atsstra{\ss}e 1, 40225 D\"usseldorf, Germany}
\author{J\"urgen Horbach}
\affiliation{Institut f\"ur Theoretische Physik II: Weiche Materie, 
Heinrich-Heine-Universit\"at D\"usseldorf, Universit\"atsstra{\ss}e 1, 40225 D\"usseldorf, Germany} 

\date{\today}% It is always \today, today,

%
\begin{abstract}
A novel thermodynamic integration (TI) scheme is presented that
allows computing the free energy of grain boundaries (GBs) in
crystals from atomistic computer simulation. Unlike previous
approaches, the method can be applied at arbitrary temperatures and
allows for a systematic extrapolation to the thermodynamic limit.
It is applied to a $\Sigma11$ GB in a face centered cubic (FCC)
Lennard-Jones crystal. At a constant density, the GB free energy
shows a non-monotonic temperature dependence with a maximum at about
half the melting temperature and the GB changes from a rigid to a
rough interface with distinct finite size scaling above this temperature.
\end{abstract}
%

%\keywords{Non-affine displacements | defects | dynamic heterogeneity | yield phenomena}
%
%
%\abbreviations{MD, Molecular Dynamics; LJ, Lenard-Jones; NAZ, non-affine zones; STZ, shear transformation zones}
\maketitle

{\it Introduction.} Most solid materials exhibit a polycrystalline
structure where crystallites of different orientation are
separated from each other by grain boundaries (GBs). The
properties of GBs in these materials determine, to a large extent,
their mechanical response and are thus of great technological
importance~\cite{Hirth1972,RobertW.Cahnc,Sutton_Balluffi_1,GBmigration,Priester,Gokhale2013}.

Many aspects of the thermodynamics and kinetics of GBs are only poorly
understood~\cite{ReadW.T.andShockley1950b,Brandon1966,Sutton1987,Gleiter1982,
Wolf1989h,Wolf1989f,Wolf1990iii,Wolf1990iv,Merkle1992,Olmsted2009b,Kokotin2014b,
Spearot2014}. In particular, there is a dearth of systematic
approaches, providing a direct calculation or measurement of
the free energy for GB formation, $\gamma$. The knowledge
of $\gamma$ would clarify the prevalence of specific GB
orientations~\cite{Saylor2003,Saylor2004b} and is essential for the
understanding of the temperature dependence of the GB's structure and
phase behavior. While at low temperatures GBs appear to be rigid, at
high temperatures GBs tend to be rough which is linked to capillary wave
fluctuations~\cite{Godreche,CL,PRIVMAN1992,PhysRevLett.112.125701}. This
change in the nature of GBs is also associated
with changes in other materials properties such as
the GB stiffness, the GB mobility, and the electrical
resistance~\cite{Hsieh1989,Kim2004,Lee2003,Lee2007,Brown2007b,Foiles2010,
Frolov2013e,divinski,Zhu2018,Straumal2012,Straumal2016,Brown2007b,PhysRevB.85.144104,BUDKE1999385,
PhysRevB.85.224107,OLMSTED20071161,Olmsted2009ii,Olmsted2011,KeGeTingsui1999}.
While the roughening transition in simple model systems such as the Ising
model has been shown to belong to the Kosterlitz-Thouless universality
class \cite{Godreche,CL,Mon1989,PRIVMAN1992}, it remains a challenge to understand
the nature of the roughening transition in GBs of metallic alloys or colloidal systems.
The GB free energy is a central quantity to reveal this issue.

GBs in metallic alloys can be analyzed down to the atomistic scale (see,
e.g., Ref.~\cite{Roesner2011}), using high-resolution transmission electron
microscopy \cite{Urban2008}.  In colloidal systems, optical microscopy
provides a particle-level view of the structure and dynamics in GBs
\cite{Gokhale2013,Lavergne2015,Lavergne2017,Edwards2014,Maire2016}.  However,
up to now, it has not been possible to determine GB free energies from
microscopy techniques. For this reason it makes it all the more important
to obtain GB free energies via particle-based computer simulations.
In this Letter, we present a novel method to achieve this goal.

There have been different attempts to determine GB (free)
energies for atomistic systems. Interfacial energy calculations
\cite{Wolf1989h,Wolf1989f,Ratanaphan2015, Kokotin2014b,Olmsted2009b}
neglect entropic contributions which are essential at finite
temperatures. Estimates in the framework of the harmonic approximation
\cite{Foiles1994,Rickman2002} are expected to work only at very
low temperatures.  However, low-temperature structures can be used
as reference states for thermodynamic integration with respect to
temperature and in fact this approach has been used in simulation studies
\cite{Broughton1986,Frolov2009,Foiles2010,Frolov2012a,Koning}. As these methods
require the interfacial free energy at a reference point, they always
involve separate harmonic or quasiharmonic calculations.  
%Besides being
%expensive this makes these methods extremely dependent on the accuracy
%of the approximate method employed and the choice of the temperature
%used as the reference. 
{\color{black}Although these methods can be used for defect free crystals \cite{Rickman2002,freitas}
or low-temperature-GBs, it is not clear how to provide a {\it
reversible} thermodynamic path around the roughening transition and the applicability of these methods to rough GBs is at least
problematic.  Another possibility to estimate the GB free energy is to use
an Einstein crystal with a GB as a reference \cite{Foiles1994}.
However, the lack of discrete translational invariance of an Einstein
crystal \cite{Frenkel1984b,polson,UMS} leads to spurious errors when
transforming to a system with rough GB and so it is also questionable
whether methods using an Einstein crystal as a reference can be employed
for rough GBs.}

The method, proposed here to calculate 
%{\color{red}\it relative} 
GB free energies 
%{\color{red}with respect to FCC}, 
from atomistic
molecular dynamics (MD) simulations, is applicable at arbitrary
temperatures below melting, as long as the crystalline order is
maintained.  It uses thermodynamic integration (TI)~\cite{UMS},
transforming a defect-free crystal into a GB structure, 
\textcolor{black}{with translationally invariant Hamiltonians throughout the TI path}.  It does not
require a low-temperature structure as a reference state and thus, both
low-temperature rigid GB boundaries as well as rough interfaces above the
roughening transition can be considered. Moreover, as we shall demonstrate
below, our TI method allows for a systematic finite-size scaling analysis
and thus the extrapolation of the finite-size GB free energies to the
thermodynamic limit. We apply our method to a Lennard-Jones system,
computing the free energy $\gamma$ of a symmetric tilt $\Sigma 11$ GB
at a constant density. As a function of temperature, $\gamma(T)$ first
increases non-linearly, followed by a monotonic decrease for temperatures
above about half the melting temperature. At high temperature, the
scaling of $\gamma$ with system size indicates that the GB is rough,
as reflected by pronounced lateral thermal undulations.

\begin{figure}
\centering
\includegraphics[width=0.45\textwidth]{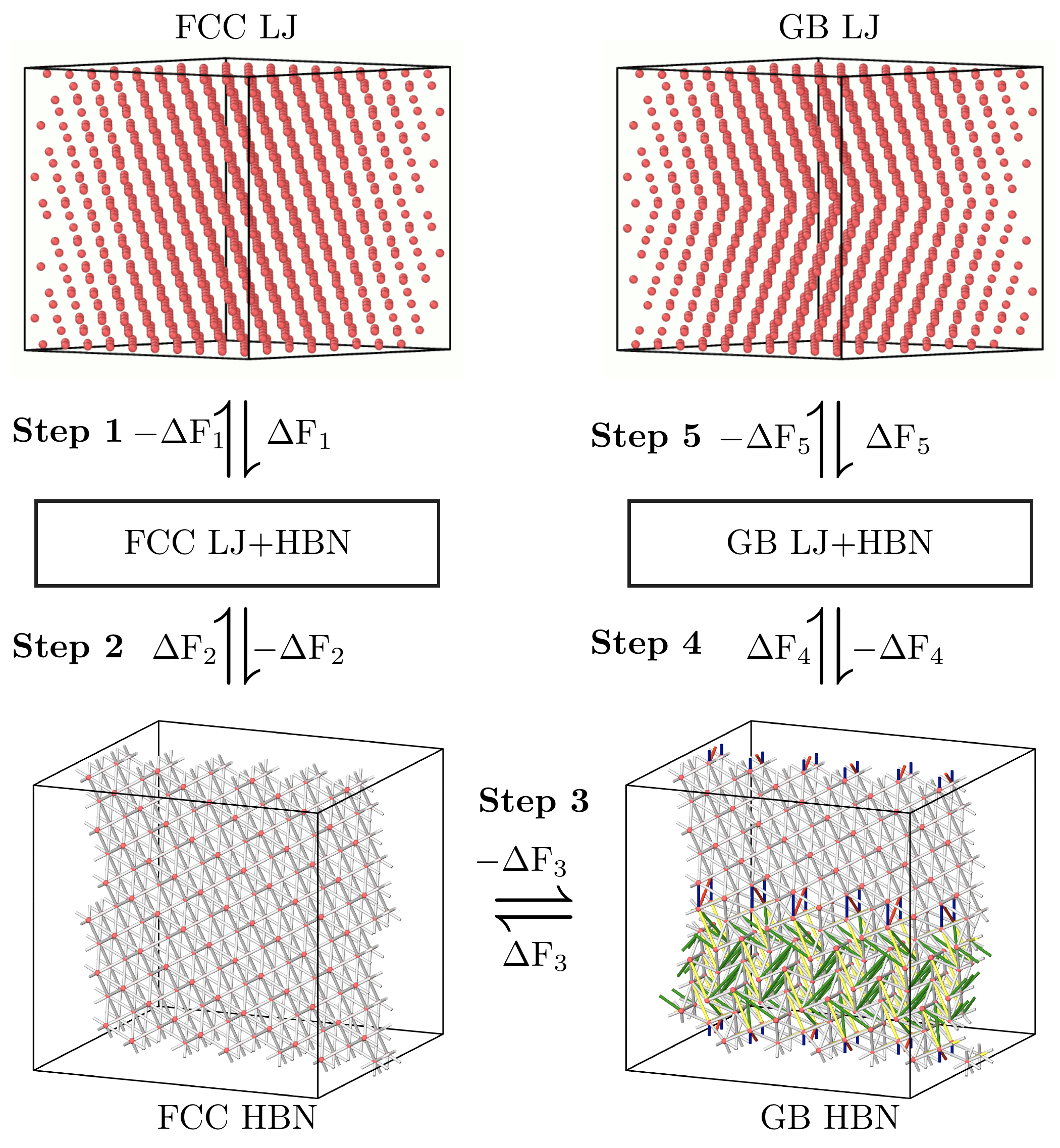}
\caption{\label{fig1} The five steps involved in the TI
path. Each of the individual steps are reversible and $\Delta
F(a\rightarrow b)=-\Delta F(b\rightarrow a)$ where $a$ and $b$ are two
states connected by a reversible TI.}
\end{figure}
\begin{figure}
\centering
\includegraphics[width=0.45\textwidth]{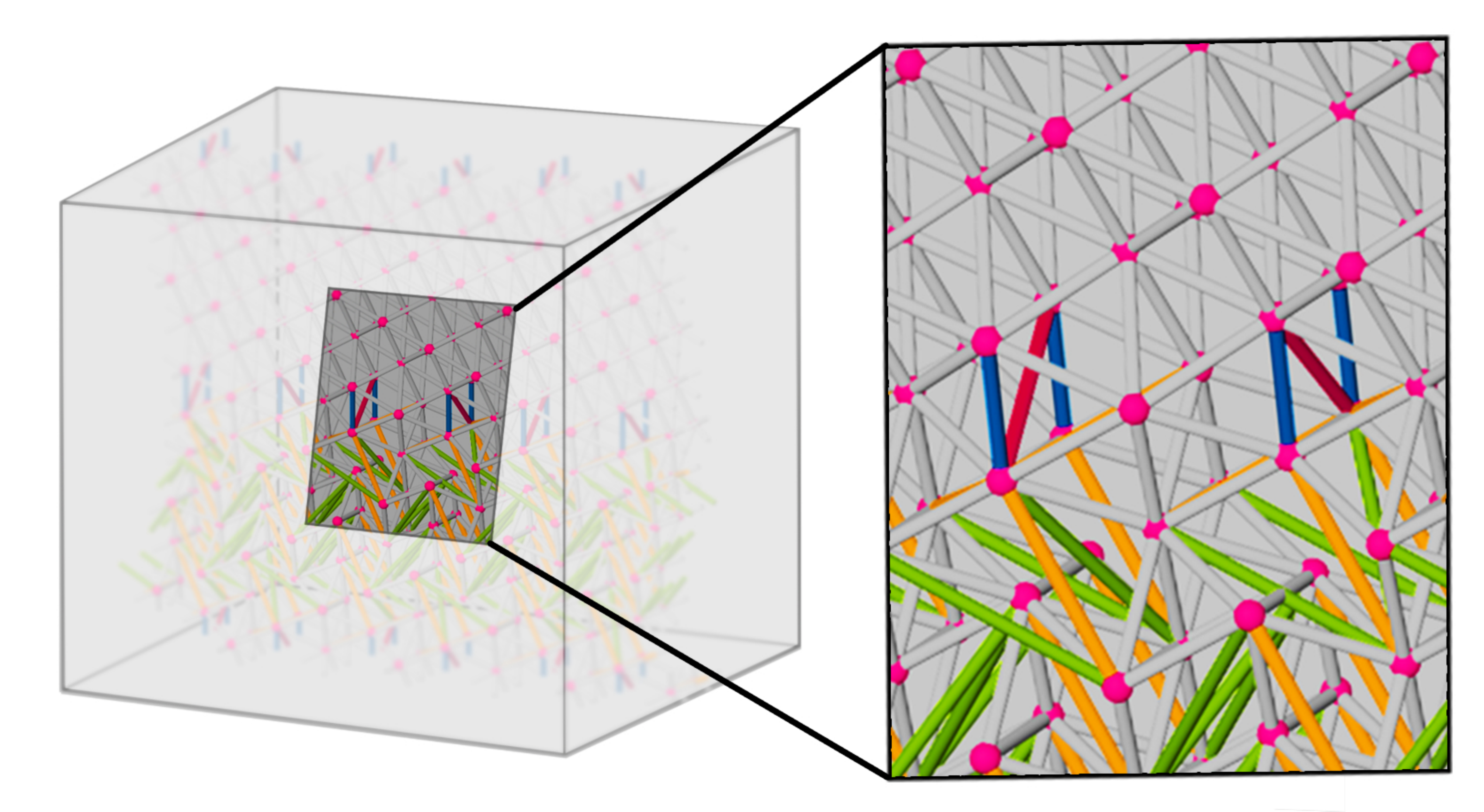}
\caption{\label{fig2} Snapshot of the HBN structure
in the GB region, as obtained at the end of step 3.  The different
colours of the bonds correspond to different bond lengths (see text).}
\end{figure}
\paragraph*{Description of the method.} The aim is to compute the free
energy cost due to the formation of a planar GB in a Lennard-Jones FCC
crystal. We consider the symmetric tilt GB $\Sigma 11(1\bar{1}3)[110]$,
placed in a simulation box with its orthogonal directions along the
vectors [3$\bar{3}\bar{2}$], [110], and [1$\bar{1}$3]. Periodic
boundary conditions are applied in all three spatial directions
and thus there are two parallel GBs, separated in $z$-direction by
$L_z/2$ and each having a total interfacial area $A=L_{x}L_{y}$. For
all the systems considered in this paper, the box dimensions $L_{x}$,
$L_{y}$, and $L_{z}$ are given in units of $l_{x}=3\sqrt{11/2}\,l$,
$l_{y}=5\sqrt{2}\,l$ and $l_{z}=2\sqrt{11}\,l$, respectively, with
$l=(4/\rho)^{1/3}$ being the lattice parameter of the FCC crystal
of density $\rho$.  The interactions between pairs of particles,
separated by a distance $r$, are modelled by the potential $u_{\rm
LJ}(r) = 4\varepsilon \left[\left(\frac{\sigma}{r}\right)^{12}
-\left(\frac{\sigma}{r}\right)^{6}\right]$, with the parameters $\sigma$
and $\varepsilon$ setting respectively the microscopic length and energy
scales of the system.  We have truncated and shifted the potential to zero
at a cut-off distance $r_{\rm cut}=2.5\,\sigma$. Reduced units are
used for all the quantities, in particular temperature and time are given
respectively in units of $\varepsilon/k_{\rm B}$ (with $k_{\rm B}=1.0$
the Boltzmann constant) and $\tau = \sqrt{m \sigma^2/\varepsilon}$
(with $m=1.0$ the mass of a particle).

The TI scheme that we propose consists of a smooth and reversible path
from a pure FCC crystal to a FCC state with two parallel GBs. From a TI
over this path, one directly obtains the GB free energy $\gamma=[F({\rm
GB})-F({\rm FCC})]/(2A)$ where $F({\rm FCC})$ corresponds to the free energy
of the initial FCC crystal and $F({\rm GB})$ to that of the final state
with two GBs.

The total transformation from the pure FCC crystal to the system with GBs
consists of five steps. In each of these steps, the system's Hamiltonian
is coupled to a parameter $\lambda$ that varies from 0 to 1. The free
energy difference between the initial state at $\lambda=0$ and the final
state at $\lambda=1$ in the $i$'th step is then given by
\begin{align}  
\label{df_ni}
\Delta F_{i} = \int_{0}^{1}
\left\langle\frac{\partial H_{i}(\lambda)}{\partial \lambda}
\right\rangle_{\lambda} d\lambda
\end{align}
with $H_i(\lambda)$ the $\lambda$-dependent Hamiltonian of the $i$'th step
and $\langle \cdots \rangle$ an ensemble or time average with respect to
the Hamiltonian $H_i(\lambda)$. Molecular Dynamics simulations
(see below) are performed at various values of $\lambda$ for
$0\le \lambda \le 1$ to obtain the integrand $\langle\frac{\partial
H_{i}(\lambda)}{\partial \lambda}\rangle$ for numerically evaluating
the integral in Eq.~(\ref{df_ni}).

The central idea of our TI scheme is to first transform the FCC crystal
with LJ interactions into a harmonic bond network (HBN) with the same
FCC structure, then transform this FCC crystal to a harmonic bond system
with GBs by introducing anisotropic equilibrium bond lengths (cf.~Figs. 1
and 2).  Finally the harmonic bond interactions are switched off and
replaced by LJ interactions between the particles, resulting in a FCC
system with GBs (see Fig.~\ref{fig1}). \textcolor{black}{The harmonic 
bond interaction potential for the FCC crystal is defined as 
$U_{\rm H}^{\rm FCC}=\frac{1}{2}k\sum_{\langle i,j\rangle}(r_{ij}-r^{\rm (eq)}_{ij})^{2}$, 
with $r_{ij}$ the instantaneous distance and $r^{\rm (eq)}_{ij}$ 
the equilibrium bond lengths between the bonded nearest neighbors 
$i$ and $j$, and $k=50$ the spring constant. The total potential energy due
to the LJ is denoted by $U_{\rm LJ}$.}

The five steps and the corresponding Hamiltonians of our TI
scheme are as follows (cf.~Fig.~\ref{fig1}): {\bf (Step 1)}
Starting with a LJ-FCC crystal, a FCC harmonic network template
(FCC HBN) is slowly turned on: $H_{1}=\textcolor{black}{E_{\rm kin}+}
U_{LJ}+\lambda^{2}U_{H}^{FCC}$ \textcolor{black}{with $E_{\rm kin}$ the
kinetic energy of the system. Here, the equilibrium bond length is set
to $r^{\rm (eq)}_{ij}=l/\sqrt{2}$.}  {\bf (Step 2)} Starting with a FCC HBN,
the LJ interactions are switched on: $H_{2}= \textcolor{black}{E_{\rm kin}+}
U_{H}^{FCC}+\lambda^{2}U_{LJ}$.  \textcolor{black}{ {\bf (Step 3)} The GB
structure with harmonic interactions (GB HBN) is transformed to a FCC
HBN with the Hamiltonian $H_{3}=E_{\rm kin}+\frac{1}{2}k\sum_{\langle
i,j\rangle}(r_{ij}-r^{\rm (eq)}_{ij})^{2}$ with a $\lambda$
dependent $r^{\rm (eq)}_{ij}=l/\sqrt{2}+(1-\lambda)^{8}c_{ij}$.}
Here, depending on the identity of the bonded particles $i$ and $j$,
$c_{ij}$ can take up values $-0.1041\,l$, $0.0\,l$, $0.2222\,l$,
$0.2929\,l$, and $0.5176\,l$, color-coded respectively as blue, grey,
red, yellow, and green bonds in Fig.~\ref{fig1} and Fig.~\ref{fig2}.
The \textcolor{black}{$r^{\rm (eq)}_{ij}$} of the grey bonds ($c_{ij}=0$)
in the GB are identical to the \textcolor{black}{$r^{\rm (eq)}_{ij}$} of the
FCC and hence they remain unchanged as $\lambda$ is changed during the
transformation of the GB to the pure FCC structure (see Fig.~\ref{fig2}
and step 3 in Fig.~\ref{fig1}).  The other bonds in the GB structure with
non-zero $c_{ij}$, however, represent bonds for which the corresponding
values of \textcolor{black}{$r^{\rm (eq)}_{ij}$} as a function of $\lambda$
are tuned such that at $\lambda=1$, they form isotropic nearest neighbor
bonds in the FCC structure while the GB structure is stabilized for
$\lambda=0$ (step 3 in Fig.~\ref{fig1}). {\bf (Step 4)} The GB HBN
is transformed to a GB structure with harmonic and LJ interactions:
$H_{4}=\textcolor{black}{E_{\rm kin}+}U_{H}^{\rm GB}+\lambda^{2}U_{\rm LJ}$
\textcolor{black}{with $U_{H}^{\rm GB}=\frac{1}{2}k\sum_{\langle i,j\rangle}
(r_{ij}-\overline{\frac{l}{\sqrt{2}}+c_{ij}})^{2}$.} {\bf
(Step 5)} A GB structure with LJ interactions is transformed to a
GB with harmonic and LJ interactions: \textcolor{black}{$H_{5}=E_{\rm
kin}+U_{LJ}+\lambda^{2}U_{H}^{\rm GB}$.}

\begin{figure}
\centering
\includegraphics[width=0.45\textwidth]{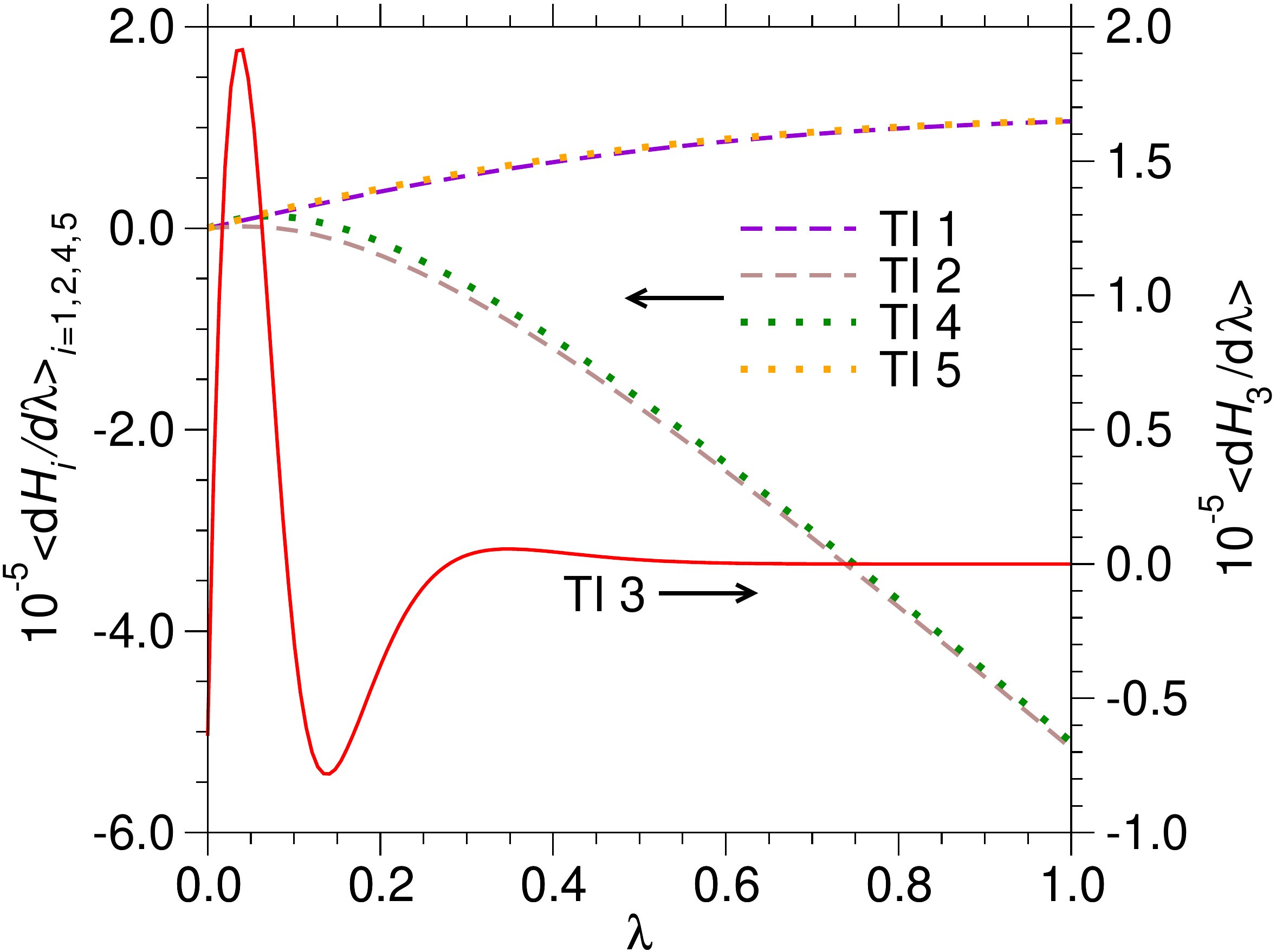}
\caption{\label{TI_Int} The integrands $\left\langle\frac{\partial
H_{i}(\lambda)}{\partial \lambda}\right\rangle$ for the five TI steps
as a function of $\lambda$ [see Eq.~(\ref{df_ni})], corresponding to a
system of $N=47520$ particles at $T=2.0$.}
\end{figure}
For the TI steps 1, 2, 4, and 5, MD simulations have been performed to
obtain the integrands in Eq.~(\ref{df_ni}) at 50 values of $\lambda$
ranging from $0$ to $1$. In the third step, the $\lambda$ axis has been
divided into $150$ equidistant intervals. As an example, the $\lambda$
dependence of the integrands for the temperature $T=2.0$ and $N=47520$
is shown in Fig.~\ref{TI_Int}. This figure indicates that the chosen
thermodynamic path leads to smooth functions for the integrands that
allow for an accurate calculation of the GB free energy $\gamma$. The
integration over these functions gives the free energy differences
$\Delta F_i$ ($i=1, \dots, 5$), corresponding to the five steps, and thus
one obtains the GB free energy as $\gamma = \frac{1}{2A} \left(\Delta
F_{1}-\Delta F_{2}-\Delta F_{3}+\Delta F_{4}-\Delta F_{5} \right)$.

\paragraph*{Simulation details.} MD simulations in the $NVT$
ensemble have been performed at density $\rho=1.179$. For all the
simulations, the systems were equilibrated for $5\times 10^{6}$ MD steps,
using a time step of $0.0001 \tau$. The integrands were
computed from the data collected at intervals of $200$ MD steps over the
subsequent $5\times 10^{6}$ MD steps. The system size analysis was done
at two temperatures $T=1.0$ and $T=2.0$ below the melting temperature,
$T_{\rm M}=2.74$~\cite{PhysRev.184.151}. To study the effect of orthogonal
distance between the GBs, system sizes $N=n\times5280$ with $n=1, 2,
\dots, 6$ were considered. For these systems the area of the GB was
kept constant with $(L_{x}, L_{y})=(2\,l_{x}, 2\,l_{y})$, while varying
$L_{z}=l_{z}, 2\,l_{z}, \dots, 6\,l_{z}$. To observe the effect
of changing $A$ on $\gamma$, three system sizes, $N=5280$, $4\times 5280$,
and $9\times 5280$, were simulated. In these cases, \textcolor{black}{we used $L_{z}=4\,l_{z}$}
and $A$ was changed with the dimensions $(L_{x},
L_{y})$ corresponding to the system sizes $(l_{x}, l_{y})$, $(2\,l_{x},
2\,l_{y})$, and $(3\,l_{x}, 3\,l_{y})$, respectively.

\begin{figure}
\centering
\includegraphics[width=0.38\textwidth]{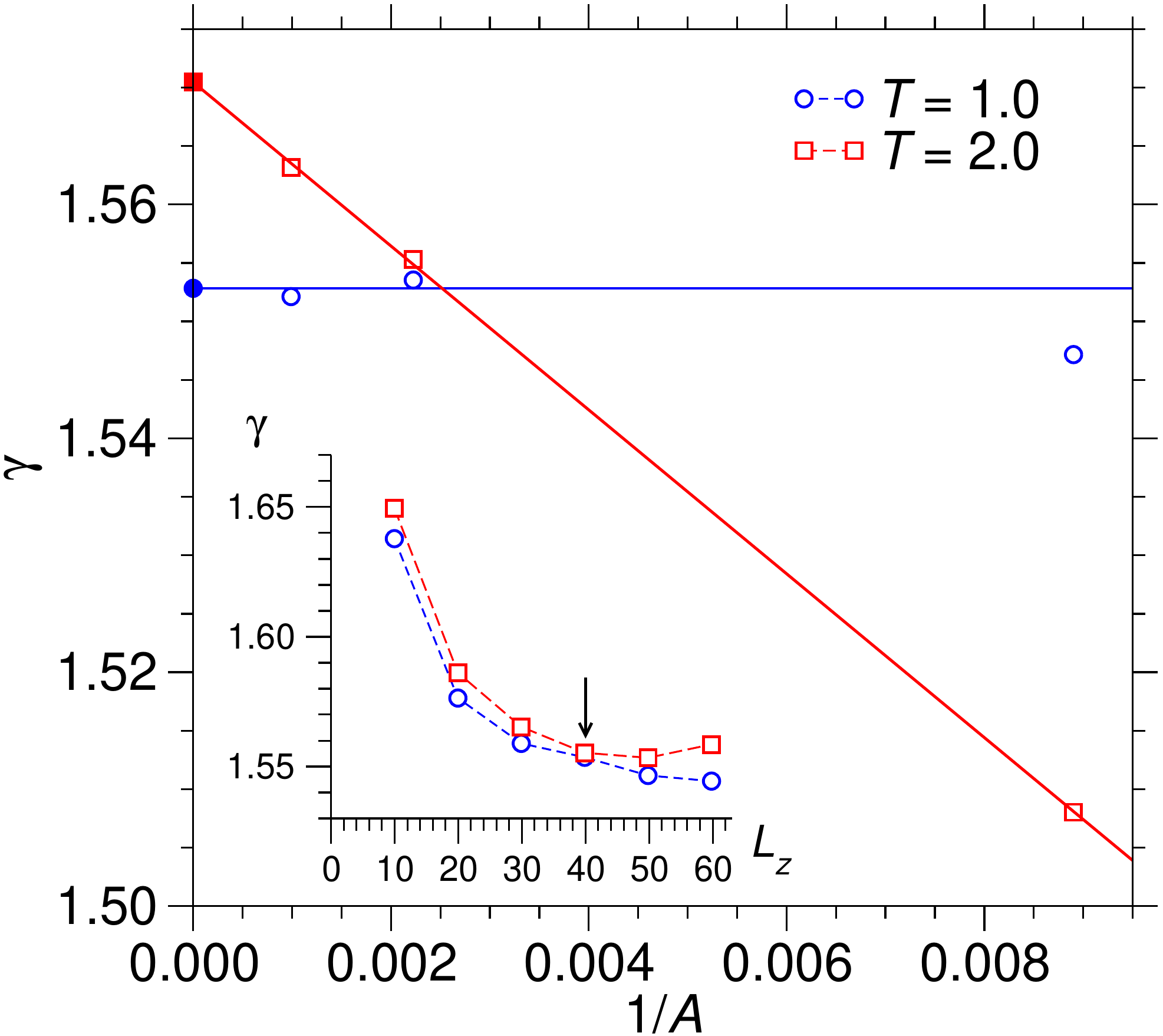}

\hspace*{0.3cm}
\includegraphics[width=0.38\textwidth]{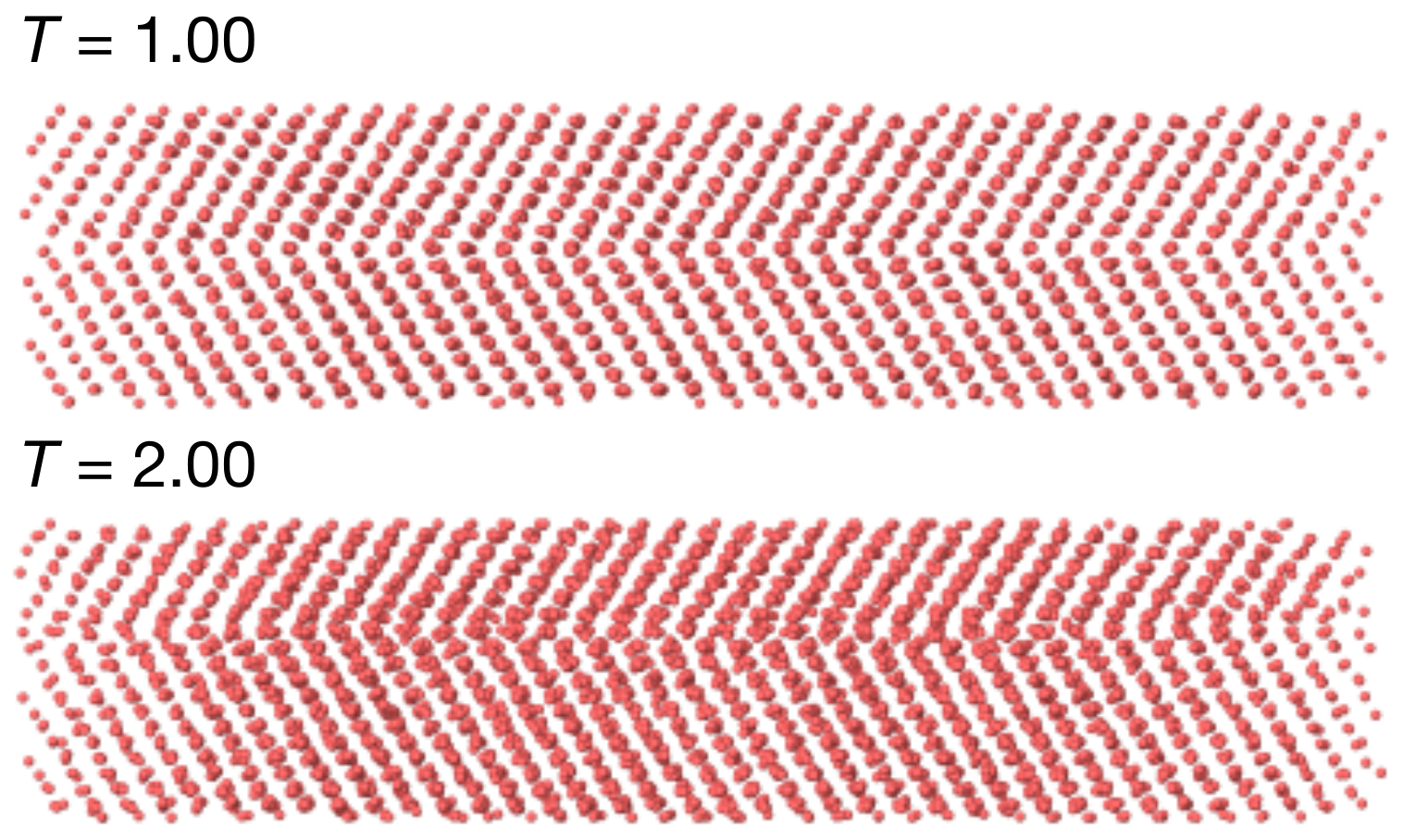}
\caption{\label{LZ_XY_F} GB free energy $\gamma$ at $T=1.0$ and $T=2.0$
as a function of $1/A$ at fixed $L_{z}\approx 40.0$ (marked by an arrow in the
inset). At $T=2.0$, a linear fit (solid red line) yields an estimate of
$\gamma$ in the thermodynamic limit (filled red square).  At $T=1.0$,
$\gamma$ saturates to a constant value (blue solid line) for $1/A
\lesssim 0.003$.  The inset displays $\gamma$ as a function of $L_{z}$,
keeping the GB interfacial area constant at $A=449.31$. The snapshots
show the GB regions of the largest systems at $T=1.0$ and $T=2.0$.}
\end{figure}
For a system of $N=47520$ with system dimensions $(L_{x}, L_{y},
L_{z})=(3\,l_{x},3\,l_{y},4\,l_{z})$, $\gamma$ has been determined at
the temperatures $T=0.0$, 0.50, 0.75, 1.00, $\dots, 2.25$.

\paragraph*{Finite-size effects.} 
%The use of periodic boundary conditions is associated with the presence of two parallel GBs orthogonal to the $z$-axis. 
If the distance between \textcolor{black}{the GB and its periodic image} is not sufficiently large,
the two GBs might ``see'' each other, thus introducing strong finite-size
corrections in our estimates of the GB free energy $\gamma$. In order
to systematically study these finite-size effects, we have computed
$\gamma$ at $T=1.0$ and $T=2.0$ for six values of the box dimension in
$z$-direction, $L_{z} = l_{z}$, $2\,l_{z}, \dots, 6\,l_{z}$, keeping
the area of the GBs constant at $A=2\,l_x\times2\,l_y=449.31$. The
results for the dependence of $\gamma$ on $L_z$ are shown in the
inset of Fig.~\ref{LZ_XY_F}. A similar behavior is found for the
two considered temperatures.  While for $L_z\lesssim 30$, $\gamma$
significantly decreases with increasing $L_z$, for larger values of $L_z$,
it essentially saturates to a constant.  Therefore, for the following
calculations we have chosen $L_{z}= 4l_{z}$ (marked with an arrow in the
inset of Fig.~\ref{LZ_XY_F}) to avoid significant finite-size corrections
due to interactions between the GBs and to investigate the effects due
to the change of the GB's area on $\gamma$.

To this end, we determine $\gamma$ for three different choices of the
GB area, $(L_x,L_y)=(l_x,l_y), (2l_x,2l_y), (3l_x,3l_y)$, where we keep
the $L_{x}:L_{y}$ ratio at approximately $1:1$. Again, the temperatures
$T=1.0$ and $T=2.0$ are considered. As can be inferred from the figure,
the scaling of $\gamma$ with $1/A$ is qualitatively different at these
temperatures. At $T=1.0$, $\gamma$ approaches the constant $\gamma=
1.554$ for sufficiently large systems (i.e.~$1/A\lesssim 0.003$), as
indicated by the horizontal line in Fig.~\ref{LZ_XY_F}.  For $T=2.0$,
however, the GB free energy shows a linear dependence on $1/A$ and the fit
with a linear function allows for an extrapolation to the thermodynamic
limit ($A\to \infty$) which gives $\gamma=1.570$ in this case (filled
square). The latter scaling behavior is expected for a rough interface,
where long-wavelength thermal undulations along the interface (capillary
waves) lead to leading-order finite-size corrections proportional
to $1/A$~\cite{PhysRevLett.112.125701,PhysRevE.90.012128}.  In fact,
the snapshots in Fig.~\ref{LZ_XY_F} indicate that the nature of the GB
changes from a rigid interface at $T=1.0$ to a rough interface at $T=2.0$.

\begin{figure}
\centering
\includegraphics[width=0.36\textwidth]{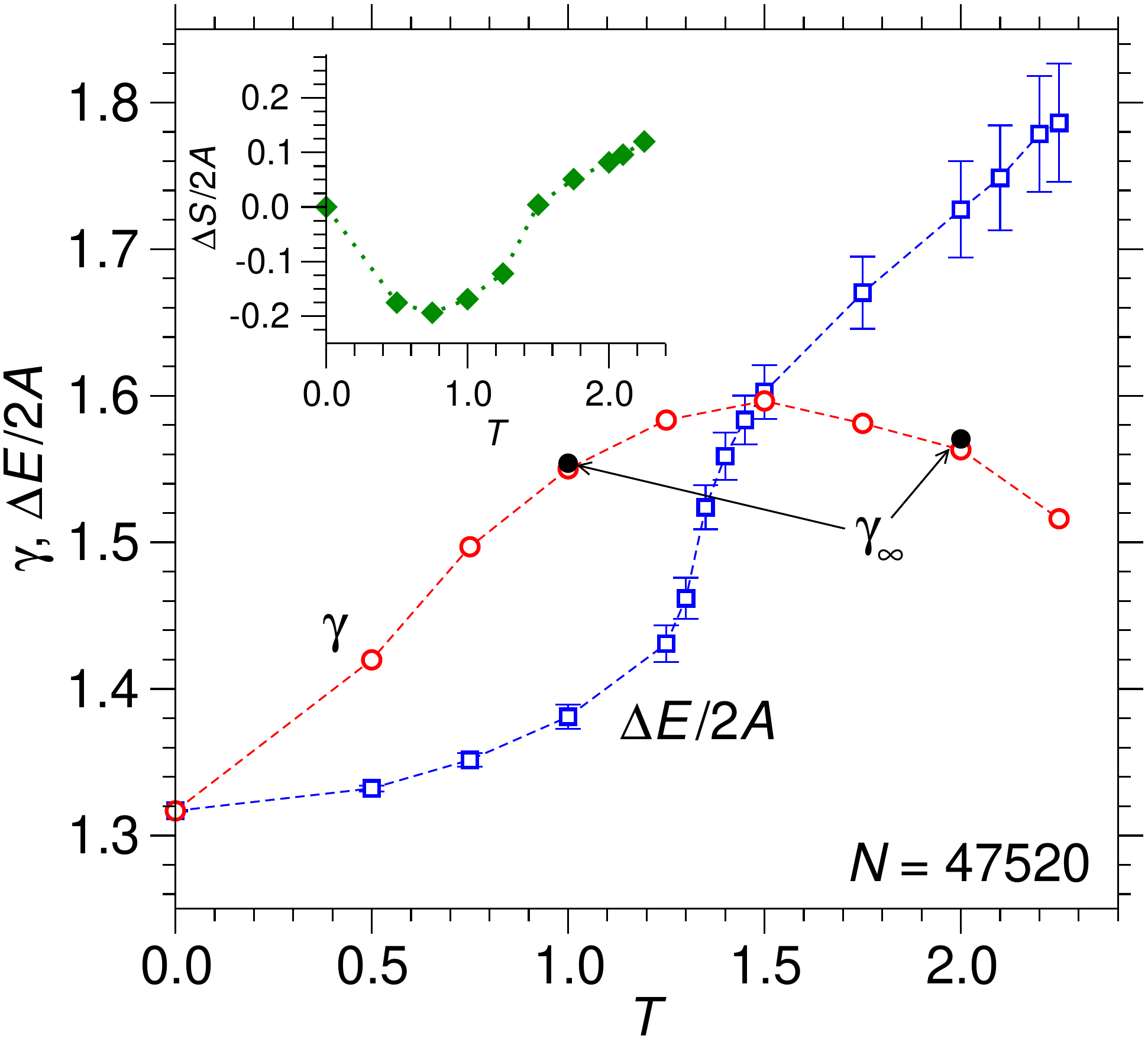}
\caption{\label{F_vs_T} GB energy $\Delta E/2A$ (blue open squares) and
free energy $\gamma$ (red open circles) as a function of temperature
for systems with $N= 47520$ particles and dimensions $(L_{x}, L_{y},
L_{z})=(3l_{x},3l_{y},4l_{z})$. The black filled circles are estimates
of $\gamma$ in the thermodynamic limit (cf.~Fig.~\ref{LZ_XY_F}). The
inset shows the excess entropy $\Delta S/2A$, see text.}
\end{figure}
\paragraph*{Temperature dependence.} Figure \ref{F_vs_T} shows the
temperature dependence of the GB free energy $\gamma$ in comparison to the
corresponding excess GB energy per unit area, $\Delta E/2A$. While the
excess energy increases monotonically with temperature, the excess free
energy $\gamma$ has a maximum at $T\approx 1.5$.  Around this maximum, the
GB undergoes a transformation from a rigid to a rough interface. Here,
the roughness is reflected in high mobility of the GB (see movies
in the supplementary material) and thermal undulations along the GB
(cf.~snapshots in Fig.~\ref{LZ_XY_F}).  Whether at high temperatures the
GBs are truly rough or whether, due to elastic effects, the capillary wave
spectrum of the thermal GB undulations is cut off at a finite wavelength,
is an open question which goes beyond the scope of the present work.

Also shown in Fig.~\ref{F_vs_T} is the temperature dependence of the GB
excess entropy per unit area, as defined by $\Delta S/(2A) = \frac{1}{T}
\left( \Delta E/(2A) - \gamma \right)$.  Below the maximum value of
$\gamma(T)$, the excess entropy is negative and it exhibits a minimum
around $T\approx 0.9$. At high temperature (where the GB is rough), it
is positive and increases linearly with temperature. How this behavior of
the excess entropy is linked to the roughening transformation of the GB,
is an open issue.

The estimates of the GB free energy in the thermodynamic limit at the
temperatures $T=1.0$ and $T=2.0$ (the two $\gamma_{\infty}$ points in
Fig.~\ref{F_vs_T}) indicate that the considered system size with $N=47520$
particles already leads to reliable values of $\gamma$ over the whole
temperature range, without being significantly affected by finite-size
effects.

We also note that the occurrence of a maximum in $\gamma(T)$ is a special
feature of the constant volume ensemble which is usually considered in
colloid experiments. Most experiments on metallic systems are performed
at constant pressure. In this case, $\gamma$ decreases monotonically with
temperature, consistent with observations in earlier simulation studies
(see supplementary material).

\paragraph*{Conclusions.} We have presented a novel TI technique to obtain
the free energy of GBs, $\gamma$, from MD simulations. The method allows
to determine $\gamma$ for both rigid and rough interfaces and to perform
a systematic finite-size scaling analysis at each temperature. While
in this work we have considered a symmetric-tilt $\Sigma 11$ GB in a
Lennard-Jones system, our TI scheme can also be applied to other types
of GBs, including asymmetric tilt GBs.  Thus, our TI technique opens the gate
to address some fundamental open questions in materials physics: What is
the nature of the interface roughening in low- and high-angle GBs \textcolor{black}{(here,
complementary information can be obtained from the determination of the
interfacial stiffness for rough GBs, as shown in Ref.~\cite{foiles_hoyt})}? What is
the prevalent GB orientation for a given material? What is the origin of
GB wetting phenomena, GB premelting, and structural transformations seen
in GBs?  These questions shall be investigated in forthcoming studies.

\begin{acknowledgments}
We thank S. Akamatsu, S. Bottin-Rousseau, J. Eiken, U. Hecht, and M. Plapp for
discussions.  The  authors  acknowledge  financial support  from  the
German  DFG  in  the  framework  of  the M-era.Net project "ANPHASES."
\end{acknowledgments}
%

%\bibliographystyle{apsrev4-1}
%\bibliography{TI_ref} 


\end{document}

\title{The free energy of grain boundaries from atomistic computer simulation : Supplementary material}

\author{Saswati Ganguly}
\affiliation{Institut f\"ur Theoretische Physik II: Weiche Materie, Heinrich 
Heine-Universit\"at D\"usseldorf, Universit\"atsstra{\ss}e 1, 40225 D\"usseldorf, Germany}
\author{J\"urgen Horbach}
\affiliation{Institut f\"ur Theoretische Physik II: Weiche Materie, Heinrich 
Heine-Universit\"at D\"usseldorf, Universit\"atsstra{\ss}e 1, 40225 D\"usseldorf, Germany} 

\date{\today}% It is always \today, today,

%

%

%\keywords{Non-affine displacements | defects | dynamic heterogeneity | yield phenomena}
%
%
%\abbreviations{MD, Molecular Dynamics; LJ, Lenard-Jones; NAZ, non-affine zones; STZ, shear transformation zones}
\maketitle

{\it TI in the isothermal-isobaric (NPT) ensemble.} The method presented in the main text to evaluate $\gamma$ of grain boundaries in the canonical (NVT) ensemble can be modified to perform the same calculation in the NPT ensemble. The thermodynamic integration steps remains identical but unlike the NVT ensemble, now the molecular dynamics simulations required to obtain the integrands in  Eq (1) should be performed at a constant temperature and pressure. 

\begin{figure}[h]
\centering
\includegraphics[width=0.36\textwidth]{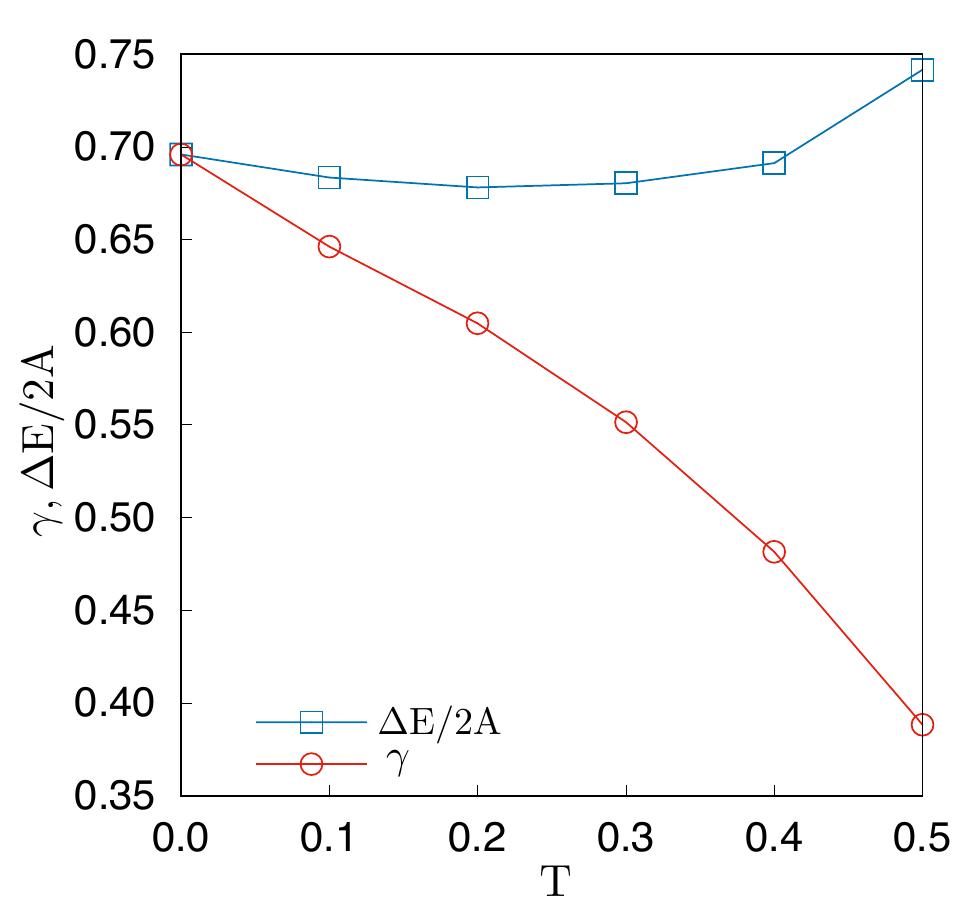}
\caption{\label{NPT_5280} Interfacial energy $\Delta E/2A$ (blue squares)
and free energy $\gamma$ (red circles) as a function of temperature
for $N= 5280$ and $P=0$.}
\end{figure}
\begin{figure}[h!]
\centering
\includegraphics[width=0.36\textwidth]{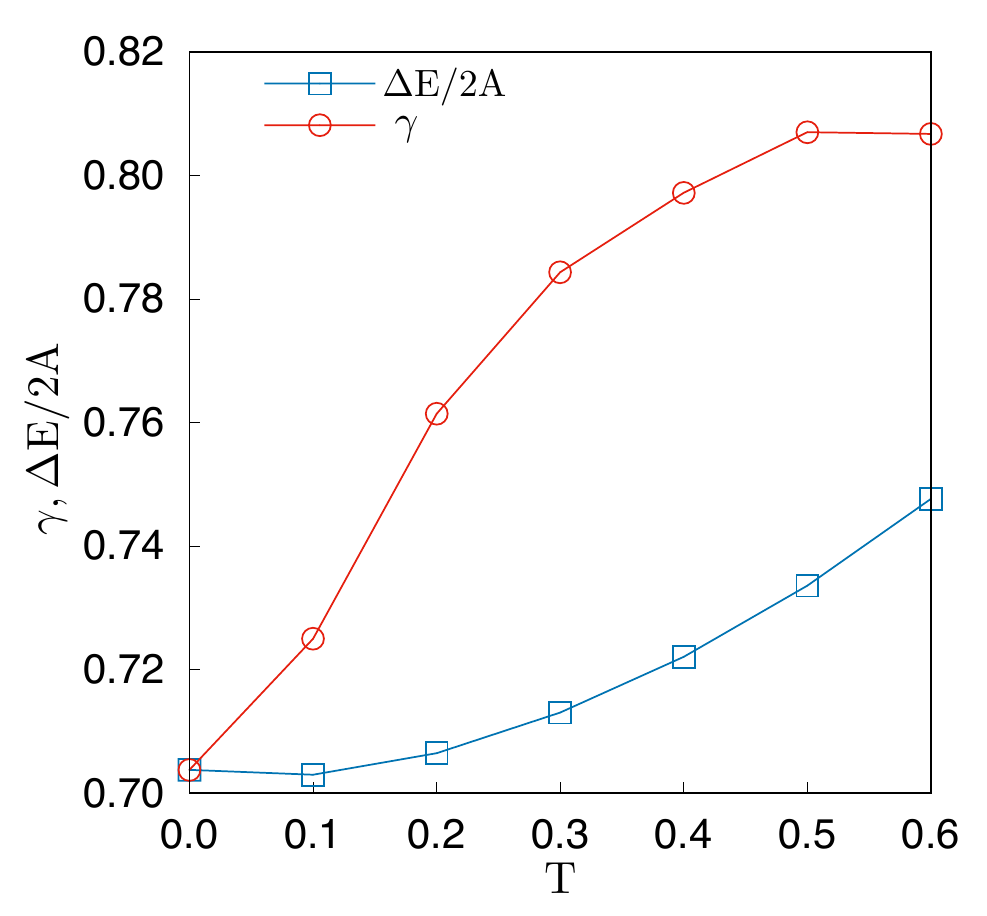}
\caption{\label{NVT_5280} Interfacial energy $\Delta E/2A$ (blue squares)
and free energy $\gamma$ (red circles) as a function of temperature
for $N= 5280$ and $\rho=1.075$.}
\end{figure}

We obtain $\gamma$ as a function of temperature while the pressure of the FCC as well as the GB structure is kept constant at P=0 for a system of $N=5280$ with dimensions $(L_{x},L_{y},L_{z})=(l_{x},l_{y},4l_{z})$. The dimensions, of course, change with the change in temperature and is determined by the density of the FCC or the GB structure at that particular temperature. Similar to the NVT ensemble, periodic boundaries are applied in all three orthogonal directions. Fig.~\ref{NPT_5280} shows $\gamma$ and the interfacial energy $\Delta E/2A$ as a function of T when the pressure is kept constant at $P=0$. The $\gamma$ shows a monotonic decrease with the increase in temperature while $\Delta E/2A$ shows relatively small changes at lower temperature with a more prominent increase at higher temperature. The temperature dependence of $\gamma$ and $\Delta E/2A$ for a system of $\rho=1.075$ and $N=5280$ in the NVT ensemble is also presented here (Fig.~\ref{NVT_5280}) for comparison. The systems chosen for the calculations in the NVT and NPT ensembles have identical configurations for the FCC crystal at $T=0$.

The method presented in this paper directly calculates interfacial free energy of a GB structure with respect to a pure FCC crystal at the same temperature.
Our results are consistent with the results obtained from previous studies~\cite{Foiles2010,PhysRevB.85.224107,Broughton1986}. Though we would like to re-iterate that $\gamma$ obtained using harmonic methods~\cite{Foiles2010} or by extrapolation to different temperatures~\cite{PhysRevB.85.224107,Broughton1986} using thermodynamic integration techniques involve very specific approximations regarding the structural and elastic properties of the GB. Despite substantial computation efforts, accuracy of these methods relies on the validity of the approximations or accuracy of the references used as starting points.

\bibliographystyle{apsrev4-1}
\bibliography{Sup_ref}